\documentclass[final]{siamltex}

\usepackage{graphicx}
\usepackage{subfigure}

\title{A novel fast solver for Poisson equation with the Neumann boundary condition\thanks{This work was supported in part by the Research Grants Council of Hong Kong (GRF 711609, 711508, 711511 and 713011), in part by the University Grants Council of Hong Kong (Contract No. AoE/P-04/08) and HKU small project funding (201007176196).}}

\author{Zu-Hui Ma\thanks{Department of Electric and Electronic Engineering, The University of Hong Kong, Pokfulam Road, Hong Kong SAR({\tt mazuhui@hku.hk}).}
        \and Weng Cho Chew\thanks{Corresponding author. Department of Electrical and Computer Engineering, University of Illinois at Urbana-Champaign (on part-time appointment with HKU), Urbana,IL 61801 USA({\tt w-chew@uiuc.edu}).}
        \and Lijun Jiang\thanks{Department of Electric and Electronic Engineering, The University of Hong Kong, Pokfulam Road, Hong Kong SAR({\tt ljiang@eee.hku.hk}).}}

\begin{document}

\maketitle

\begin{abstract}
In this paper we present a novel fast method to solve Poisson equation in an arbitrary two dimensional region with Neumann boundary condition. The basic idea is to solve the original Poisson problem by a two-step procedure: the first one finds the electric displacement field $\mathbf{D}$ and the second one involves the solution of potential $\phi$. The first step exploits loop-tree decomposition technique that has been applied widely in integral equations within the computational electromagnetics (CEM) community. We expand the electric displacement field in terms of a tree basis. Then, coefficients of the tree basis can be found by the fast tree solver in $O(N)$ operations. Such obtained solution, however, fails to expand the exact field because the tree basis is not completely curl free. Despite of this, the accurate field could be retrieved by carrying out a procedure of divergence free field removal. Subsequently, potential distribution $\phi$ can be found rapidly at the second stage with another fast approach of $O(N)$ complexity. As a result, the method dramatically reduces solution time comparing with traditional FEM with iterative method. Numerical examples including electrostatic simulations are presented to demonstrate the efficiency of the proposed method.
\end{abstract}

\begin{keywords}
fast Poisson solver, loop-tree decomposition, electrostatic problems
\end{keywords}

\begin{AMS}
15A15, 15A09, 15A23
\end{AMS}

\pagestyle{myheadings}
\thispagestyle{plain}
\markboth{A NOVEL FAST SOLVER FOR POISSON EQUATION}{ZU-HUI MA, WENG CHO CHEW AND LIJUN JIANG}

\section{Introduction}
There are a variety of physical situations and engineering problems described by elliptic partial differential equations (PDEs) such as the Poisson equation:
$$
\nabla^2 u(\mathbf{r})=f(\mathbf{r}) .
$$
Examples of this equation are encountered in low-frequency dielectric or conductivity problems \cite{Jackson_EM_book}\cite{Barkas}.   This equation is often solved in micro and nanoelectronic device physics: it is also found in electronic transport and electrochemistry in terms of the Poisson-Boltzmann equation \cite{Pois_BoltzEqu}\cite{Adelmann2010}.  Hence, finding a robust and efficient way to solve it has attracted considerable interests in various fields of research.

Over the past few decades, several kinds of fast methods for solving Poisson equation have been proposed. One popular fast Poisson solver is based on Fourier analysis and accelerated by FFT \cite{Barkas}. However, this method has generally been limited to regular geometries, such as rectangular regions, 2D polar and spherical geometries \cite{LaiPS2002}, and spherical shells \cite{HuangFFT2011}.

Multigrid methods are generally accepted as among the fastest numerical methods %\cite{MG_book}\cite{MG_rev}\cite{MGtutorial2000}%
[7-9]. These methods take advantage of fine mesh and coarse mesh. Multigrid methods can be used in irregular domains and also be extended to other partial differential equations. But it is difficult to implement multigrid methods in a robust fashion because they demand a hierarchy of grids of different density, which are not convenient in many real world problems.

Another algorithm known as the fast multipole method (FMM) %\cite{McKenney1995}\cite{Ethridge2001}\cite{Langston2011}
 [10-12] has been developed to solve Poisson problems with $O(N\log N)$ complexity.  This method is applied to integral equation derived via the Green's function rather than differential methods where Poisson equation is discritized directly. FMM solvers are particularly well suited for solving irregular shape problems.

In this paper, we will solve Poisson equation with Neumann boundary condition, which is often encountered in electrostatic problems, through a newly proposed fast method. Instead of discretizing Poisson equation directly, we solve it in two sequential steps: The first step aims to find the displacement electric field $\mathbf{D}$ from $\nabla\cdot\textbf{D}=\rho$, where $\rho$ is the charge distribution; then, the second one obtains potential $\phi$ from $\nabla\phi=-\textbf{D}/\epsilon$, in which $\epsilon$ stands for the permittivity of the medium.

The remainder of the paper is organized as follows. In Section \ref{sec:PD}, we outline the problem arising from electrostatics that we seek to solve. In Section \ref{sec:SM}, details of the proposed method will be described, in which two sequential steps of this method will be presented by two subsections. Next, in Section \ref{sec:NE}, we demonstrate and validate the efficiency by applying several numerical examples. Finally, the conclusions are drawn in Section \ref{sec:CL}.

\section{Problem Description}
\label{sec:PD}

The general Poisson equation for heterogeneous media is expressed as:
\begin{equation}
\nabla\cdot\epsilon_r(\mathbf{r})\nabla\phi(\mathbf{r})=-\frac{\rho(\mathbf{r})}{\epsilon_0}
\end{equation}
where $\phi(\mathbf{r})$ is the potential and $\rho(\mathbf{r})$ describes the charge density, $\epsilon_r(\mathbf{r})$ and $\epsilon_0$ are relative permittivity and free space permittivity, respectively.

\begin{figure}[t]
\centerline{\includegraphics[width=0.4\columnwidth,draft=false]{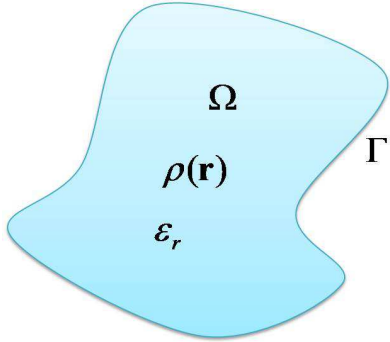}}
\caption{Schema of the electrostatic problem}
\label{fig:test}
\end{figure}

In this paper, we focus on this kind of Poisson problems, as illustrated in Fig.~\ref{fig:test}, where the charge density $\rho(x,y)$ distributes within a two dimensional domain $\Omega$ and the relative permittivity $\epsilon_r$ is a constant. Furthermore, the Neumann boundary condition is imposed on $\Gamma$.

%\begin{figure}[t]
%\vspace{3in}
%\caption{{\rm Log}$_{10}$ of the residual norm versus the number of
%{\rm GMRES}$(m)$ iterations for $c=d=10$ with fast Poisson
%preconditioning. Solid curve: Algorithm {\rm EHA}; dotted
%curve: {\rm FDP} method; dashed curve: {\rm FSP} method.}
%\label{pdep1}
%\end{figure}

Thus the equation to be solved is
\begin{eqnarray}\label{PE}
\nabla^2\phi(x,y)=-\frac{\rho(x,y)}{\epsilon_r\epsilon_0} \mbox{ }\mbox{ }\mbox{ }\mbox{ }\mbox{ for $(x,y)\in\Omega$} \nonumber  \\
\frac{\partial\phi(x,y)}{\partial n}=g(x,y) \mbox{ }\mbox{ }\mbox{ }\mbox{ }\mbox{ }\mbox{ }\mbox{ }\mbox{ for $(x,y)\in\Gamma$}
.
\end{eqnarray}

\section{Solution method}
\label{sec:SM}

The proposed new method solves (\ref{PE}) by two steps. More specifically, we first solve this equation
\begin{equation}
\nabla\cdot\textbf{D}=\rho(x,y)
\end{equation}
where the electric displacement field $\textbf{D}$ is the product of permittivity $\epsilon$ and electric field $\textbf{E}$, namely
\begin{equation}
\textbf{D}=\epsilon\textbf{E}
\end{equation}
and
\begin{equation}
\epsilon=\epsilon_r\epsilon_0.
\end{equation}
Once we have the electric displacement field, the potential distribution $\phi$ can be found by solving
\begin{equation}
\nabla\phi(x,y)=-\frac{\textbf{D}}{\epsilon},
\end{equation}
which corresponds to the second step.

Although two steps are required for this algorithm, fast methods can be applied to both steps by virtue of the tree basis.

\subsection{Solution of the Electric Field}\label{subSec:1St}
In order to find the electric field distribution, the problem at this stage is represented by
\begin{eqnarray}\label{s1e}
\nabla\cdot\textbf{D}=\rho(x,y)\mbox{ }\mbox{ }\mbox{ }\mbox{ }\mbox{ for $(x,y)\in\Omega$} \nonumber  \\
\hat{n}\cdot\textbf{D}=\frac{g(x,y)}{\epsilon_r\epsilon_0} \mbox{ }\mbox{ }\mbox{ }\mbox{ }\mbox{ for $(x,y)\in\Gamma$}
\end{eqnarray}
and
\begin{equation}
\nabla\times\textbf{E}=0.
\end{equation}

For the homogeneous medium, we can further infer that the electric displacement field is curl free as well ($\nabla\times\textbf{D}=0$), since the relative permittivity $\epsilon_r$ is constant.

\subsubsection{Construct Matrix System by the Tree Basis}

In computational electromagnetics community, the low frequency problem has attracted intensive research for the last decade. When the frequency is low, the current $\textbf{J}$ naturally decomposes into a solenoidal (divergence free) part and an irrotational (curl free) part which is known as the Helmholtz decomposition \cite{ChewBook2008}. Moreover, these two parts are not balanced when the frequency becomes low. Hence, there is a severe numerical problem when solving integral equations in which RWG bases are normally used. One remarkable remedy to this low frequency breakdown is the well known loop-tree decomposition %\cite{Wilton1981}\cite{Mautz1984}\cite{JSZhao2000}\cite{Wu1994}\cite{Burton1995}
[14-18]. The RWG basis set is decomposed into the loop basis, which has zero divergence, and the tree basis, which has non-zero divergence. This is a quasi-Helmholtz decomposition because the tree basis is not curl free.

Borrowing the idea of loop-tree decomposition, we can use a tree basis to expand the irrotational electric displacement field as
\begin{equation}\label{Dtree}
\textbf{D}_{tree}=\sum_{i=1}^{N_t} t_i\textbf{T}_i(x,y)
\end{equation}
where $\textbf{T}_i$ is the $i$-th tree basis whose coefficient is $t_i$, and $N_t$ is the total number of tree basis functions. It is equal to the number of patches minus one for a triangular mesh.
We then use the pulse basis to expand the charge density $\rho$, namely
\begin{equation}\label{charge}
\rho(x,y)=\sum_{i=1}^{N_p} c_i P_i(x,y)
\end{equation}
where $N_p$ stands for the patch number. The pulse function is defined as
$$
P_i(x,y)=\left\{ \begin{array}{rl}
 1 &\mbox{ if $(x,y)\in$ \textit{i}-th patch } \\
  0 &\mbox{ otherwise}
       \end{array} \right.
$$

Substituting (\ref{Dtree}) and (\ref{charge}) into (\ref{s1e}) and testing it with the same set of pulse functions as in the Galerkin's method, we have a matrix equation
\begin{equation}\label{S1S}
\bf{\overline{K}}\cdot\textbf{I}_t= \textbf{V}_{\rho}
\end{equation}
where
\begin{equation}
\left[\bf{\overline{K}}\right]_{ij}= \int\int P_i(x,y)\nabla \cdot \textbf{T}_j(x,y)\,\mathrm{d}x\mathrm{d}y
\end{equation}
\begin{equation}
\left[\textbf{V}_{\rho}\right]_i= \int\int P_i(x,y)\rho(x,y)\,\mathrm{d}x\mathrm{d}y = c_i
\end{equation}

\subsubsection{Fast Tree Solver}

\begin{figure}[t]
\centerline{\includegraphics[width=0.6\columnwidth,draft=false]{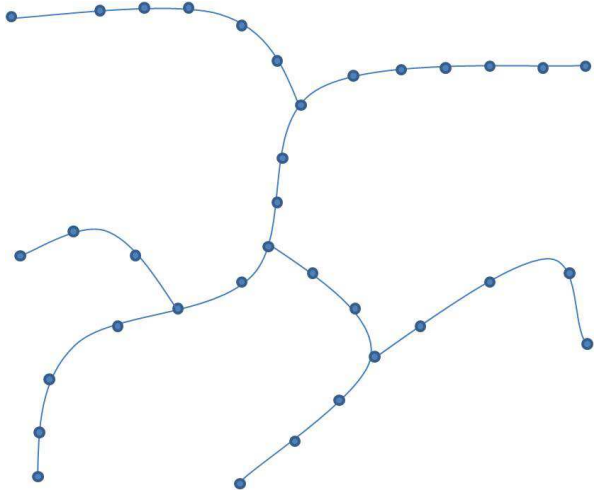}}
\caption{Schematic of a general tree}
\label{fig:tree}
\end{figure}

As described in \cite{ChewBook2008}\cite{JSZhao2000}, (\ref{S1S}) can be solved in $O(N)$ operations. We outline the basic idea as follows. In the first place, a tree is found to span the surface where Poisson equation is solved. This tree corresponds to a set of RWG basis functions complementary to the loop basis. Fig.\mbox{ }\ref{fig:tree} shows a typical tree structure. Each node of the tree represents the center of a patch, which is related to the triangle patch of the RWG basis. Each line between two nodes can be associated with the current that flows between two patches. Furthermore, the patch unknown associated with a tip of the dendritic branch is only related to one current unknown. Therefore, starting from the branch tips, the unknowns can be solved for recursively until a junction is reached. Noticing the fact that a junction node cannot connect more than three neighboring nodes for the case of RWG functions, we can solve for the unknowns of junctions when the unknowns on two associated open branches have been solved. Hence, all unknowns can be solved for in $O(N)$ operations.

\subsubsection{Divergence Free Field Removal}

It is well known that the tree basis is not curl free. This property makes the $\textbf{D}_{tree}$ from (\ref{S1S}) inaccurate because some solenoidal components contaminate it. Moreover, the tree basis is non-unique. Theoretically, the $\textbf{D}_{tree}$ is a vector living in the space spanned by the RWG basis, which satisfies Eq. (\ref{s1e}). From this point of view, the desired $\textbf{D}$ can be obtained by projecting $\textbf{D}_{tree}$ onto the curl free space. However, it is formidable to find a basis set which is completely curl free in the RWG space. In \cite{MR2005}, SVD method was used for finding the solenoidal basis set. But SVD is expensive in terms of both memory and computing time requirements.

Another way to circumvent this obstacle is to remove the complementary solenoidal (divergence free) part that is orthogonal to the curl free space. This is practical because the divergence free part, unlike its curl free counterpart, could be expanded by the loop basis that is living on the space spanned by RWG functions. It is noted that the loop basis has divergence free property; hence, it can expand the divergence free space completely. Therefore, we can first use a loop basis to project $\textbf{D}_{tree}$ onto the divergence free space. Once the projection has been done, the pure curl free part of $\textbf{D}_{tree}$ can be obtained by subtracting the divergence free components.

Mathematically, $\textbf{D}_{tree}$ can be expressed by
\begin{equation}\label{Dtl}
\textbf{D}_{tree}= \sum_{i=1}^{N} a_i\textbf{f}_i(x,y)+\sum_{i=1}^{N_l} l_i\textbf{L}_i(x,y)
\end{equation}
where the first term on right side corresponds to pure curl free part of $\textbf{D}_{tree}$, while the second one associates the divergence free part. Moreover, $\textbf{f}_i$ refers to RWG basis functions whose total number is $N$, and $\textbf{L}_i$ denotes the loop basis functions with total number of $N_l$.

By using the loop basis to test the above equation, Eq.\mbox{ }(\ref{Dtl}) leads to a matrix system
\begin{equation}\label{PrjS}
\overline{\bf G}_l \cdot \textbf{I}_l= \textbf{V}_d
\end{equation}
where
\begin{equation}
\left[\overline{\bf G}_l \right]_{ij}=\int\int \textbf{L}_i(x,y) \cdot \textbf{L}_j(x,y)\,\mathrm{d}x\mathrm{d}y
\end{equation}
is the Gram matrix of loop basis, and
\begin{equation}
\left[ \textbf{V}_d \right]_i=\int\int \textbf{L}_i(x,y) \cdot \textbf{D}_{tree}(x,y)\,\mathrm{d}x\mathrm{d}y
\end{equation}
is the projection value of $\mathbf{D}_{tree}$ onto the loop space. In this equation, we use the relation
\begin{equation}
\int\int \textbf{L}_i(x,y) \cdot \sum_{i=1}^{N} a_i\textbf{f}_i(x,y) \,\mathrm{d}x\mathrm{d}y=0
\end{equation}
because the fact that loop basis is orthogonal to bases of curl free space.

The Gram matrix of loop bases, $\overline{\bf G}_l$,  is highly sparse. Moreover, it is a symmetric, positive definite and diagonal dominant matrix. Therefore, common iterators, such as CG, Bi-CGSTAB \cite{BiCGSTAB} and GMRES \cite{SaadBook}\cite{gmres1986}, work efficiently for solving Eq.\mbox{ }(\ref{PrjS}).

Consequently, the electric displacement field desired can be retrieved by
\begin{equation}
\textbf{D}=\textbf{D}_{tree}-\sum_{i=1}^{N_l} l_i\textbf{L}_i
\end{equation}

\subsection{Solution of Potential}
In the second stage, we need to solve the following equation to find potential
\begin{equation}
\nabla\phi(x,y)=-\frac{\textbf{D}}{\epsilon}.
\end{equation}
This can be achieved by using the same method of the above section because the del operator ($\nabla$) is the transpose of the divergence operator ($\nabla\cdot$).
Hence, we expand the potential $\phi$ in terms of pulse functions, that is,
\begin{equation}\label{Phi}
\phi(x,y)=\sum_{i=1}^{N_p}\nu_i P_i(x,y)
\end{equation}
With the similar method described in Section \ref{subSec:1St}, testing this equation using a set of tree basis, we obtain
\begin{equation}\label{S2S}
\overline{\bf K}^T \cdot \textbf{I}_\phi= \textbf{V}_\phi
\end{equation}
where
\begin{equation}
\left[\textbf{V}_\phi \right]_{i}=-\int\int \textbf{T}_i(x,y) \cdot \frac{\textbf{D}(x,y)}{\epsilon_r \epsilon_0}\,\mathrm{d}x\mathrm{d}y
\end{equation}
and $\left[\textbf{I}_\phi \right]_{i}=\nu_i $ which is defined in (\ref{Phi}). The element $\overline{\bf K}^T$ is originally defined as
$$
\left[\overline{\bf K}^T \right]_{ij}=-\int\int \textbf{T}_i(x,y) \cdot \nabla P_j(x,y)\,\mathrm{d}x\mathrm{d}y ,
$$ which can be deduced using integration by parts as
\begin{equation}
\left[\overline{\bf K}^T \right]_{ij}=\int\int  P_j(x,y)\nabla \cdot \textbf{T}_i(x,y)\,\mathrm{d}x\mathrm{d}y=\left[\overline{\bf K} \right]_{ji}
\end{equation}

Consequently, the resulting matrix is just a transpose of the matrix $\left[\overline{\bf K} \right]$ from Section \ref{subSec:1St}. Hence, (\ref{S2S}) can be solved with fast tree solver in $O(N)$ operations.

\section{Numerical Examples}
\label{sec:NE}
In this section, two numerical examples are shown to validate the efficiency of the proposed method. All examples have been calculated on a standard computer with $2.66$ GHz CPU, $4$ GB memory and Windows operating system.

\subsection{ Simple Neumann Problem}\label{sSec:SP}
In order to validate the correctness of this algorithm, we solve a simple 2-D Poisson equation as the first example. The Poisson equation in this case is
\begin{equation}
\frac{\partial^2 \phi}{\partial x^2}+\frac{\partial^2 \phi}{\partial y^2}=-\pi \cos(\pi x)-\pi \cos(\pi y)
\end{equation}
where $(x,y)\in \Omega = \left[ 0,1 \right] \times \left[ 0,1 \right]$, with the Neumann boundary condition
\begin{equation}
\hat{\bf n} \cdot \left(\hat{\bf x} \frac{\partial \phi}{\partial x}+\hat{\bf y} \frac{\partial \phi}{\partial y} \right)=0 \mbox{~~~~~~~~~$(x,y) \in \Gamma$.}
\end{equation}

It is well known that Neumann problems is uniquely solvable but up to a constant; hence,we have to impose a reference potential so that the uniqueness can be guaranteed. By setting the potential as $2/\pi$ at the point $(0,0)$, the  aforementioned problem has an analytical solution
\begin{equation}
\phi(x,y)= \left[\cos(\pi x)+ \cos(\pi y)\right]/ \pi.
\end{equation}

\begin{figure}[t]
%\centering
\subfigure[Simulation potential distribution]{
\includegraphics[scale=0.3]{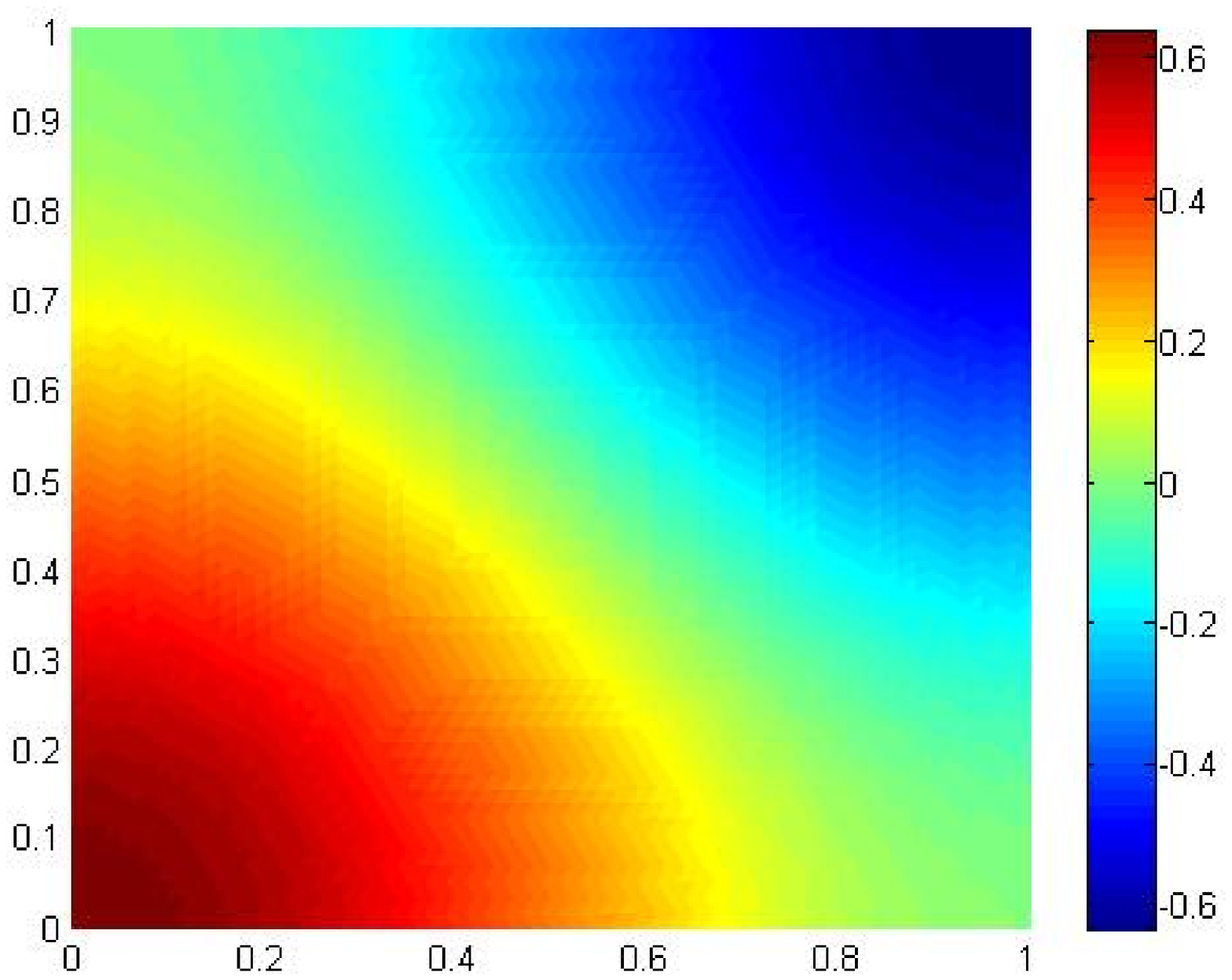}
\label{fig:subfig1}
}
\subfigure[Exact potential distribution]{
\includegraphics[scale=0.3]{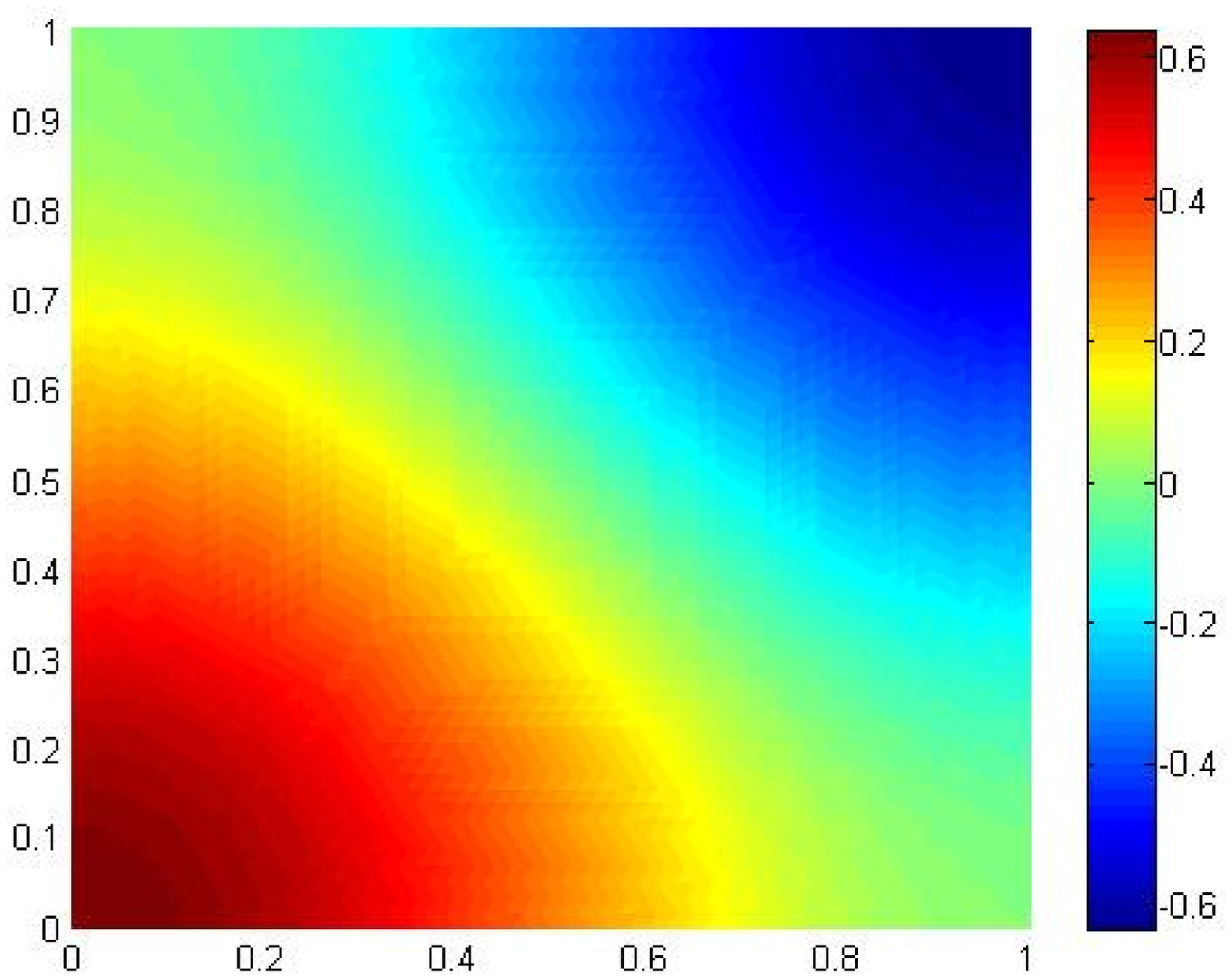}
\label{fig:subfig2}
}
\caption[Optional caption for list of figures]{Potential distribution:\subref{fig:subfig1} Simulation result,\subref{fig:subfig2} Exact potential distribution}
\label{fig:Test1}
\end{figure}

\begin{figure}[t]
\centerline{\includegraphics[width=8cm,draft=false]{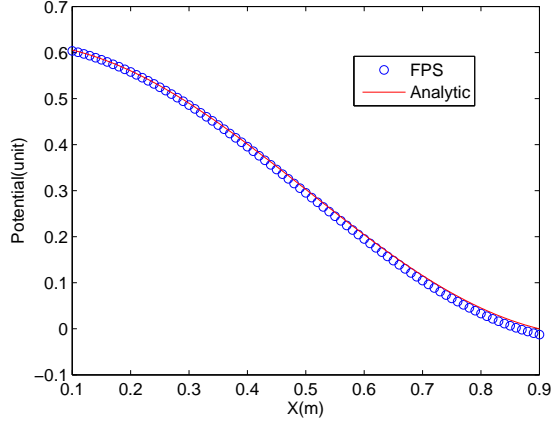}}
\caption{Comparisons of values along one line ($y=0.1$)}
\label{cprPot1}
\end{figure}

By utilizing the proposed method, potential distribution can be found as shown in Fig.\mbox{ }\ref{fig:Test1}, where Fig.\mbox{ }\ref{fig:subfig1} shows the potential distribution result while Fig.\mbox{ }\ref{fig:subfig2} gives the exact solution as a reference. Furthermore, in Fig.\mbox{ }\ref{cprPot1}, we compared the value from the proposed approach with the analytic results. It is obvious that the potential distribution from this new method agrees with analytical one well.

\subsection{Line Source Problem}

\begin{figure}[t]
\centerline{\includegraphics[width=9cm,draft=false]{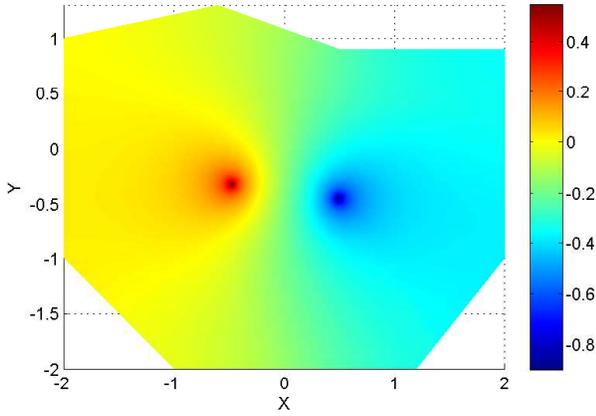}}
\caption{Potential distribution calculated by the proposed method for line sources}
\label{ellDip}
\end{figure}

As the second example, a Poisson problem involving two line charge sources was simulated within a polygonal region. One line source was at point $(-0.475,-0.329)$ with a positive unit charge while the other one was at point $(0.494,-0.459)$  with a negative unit charge. The resultant potential distribution is shown in Fig. \ref{ellDip}. In this case, the potential value of the up-left corner $(-2.0,1.0)$  was set to be zero as a reference potential. The result has been validated by FEM method.

\subsection{Computational Complexity Analysis}

To analyze the computational complexity, we have computed the problem of Section \ref{sSec:SP} with different mesh densities. The solution time of this algorithm consists of three parts: the first fast tree solution time for (\ref{S1S}), the divergence free field removal time and the second fast tree solution time for (\ref{S2S}). As for the fast tree solution time, Zhao and Chew have proved that it is of $O(N)$ complexity \cite{JSZhao2000}. Hence, the bottleneck comes from the procedure of divergence free field removal, which amounts to the iterative solution time of (\ref{PrjS}). The CPU time of this part therefore depends on iterative solver type and the required accuracy level.

In divergence free field removal part, the Gram matrix of loop basis has the form
\begin{equation}\label{GramL}
\left[\overline{\bf G}_l \right]_{ij}=\int\int \textbf{L}_i(x,y) \cdot \textbf{L}_j(x,y)\,\mathrm{d}x\mathrm{d}y.
\end{equation}
As similar to the stiffness matrix of FEM, this matrix is a symmetric, positive definite system. For this kind of matrix systems, the number of iterations is proportional to the square-root of the condition number.

\begin{figure}[t]
\centerline{\includegraphics[width=8cm,draft=false]{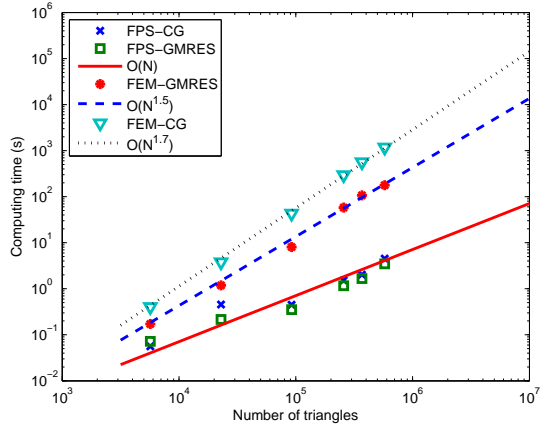}}
\caption{Complexity comparison with same stopping criterion}
\label{cplx}
\end{figure}

Fig.~\ref{cplx} shows the computational complexity comparison between this fast Poisson solver and FEM. Both conjugated gradient (CG) and GMRES are adopted with same stopping criterion ($0.01$). For GMRES, the restart parameter is $60$.  As can be seen from this plot, the total solution time of our proposed new method is evidently less than the one of traditional FEM, whichever CG or GMRES is adopted. In our numerical experiment, the complexity of this new method approaches linear complexity. Moreover, GMRES methods work better than CG method in our simulations.

%Fig.~\ref{cplx} shows the total CPU time of solution procedures when the conjugate gradient (CG) iterative method is used and the iteration number is fixed. As can be seen from this plot, the total solution time is close to linear complexity. However, the iteration number increases with the patch number if we want more accurate results. This could affect the overall computational complexity to increase to more than $O(N)$. Relevant research is being conducted to reduce its influence.

\section{Conclusions}
\label{sec:CL}
We have proposed and developed a new two dimensional fast Poisson solver for electrostatic problems with the Neumann boundary condition based on the loop-tree decomposition technique. This method could solve Poisson equation rapidly with the fast tree solver. Its computational consumption is evidently less than the one of traditional FEM. This method promises to be a novel fast Poisson solver that can solve general Poisson equation for electrostatics as well as other related fields.

\section*{Acknowledgments}
The authors would like to thank Dr. Sheng Sun, Dr. Min Tang and Jun Huang for their helpful discussions and feedback from Leslie Greengard.

%This work was supported in part by the Research Grants Council of Hong Kong (GRF 711609, 711508, 711511 and 713011), in part by the University Grants Council of Hong Kong (Contract No. AoE/P-04/08) and HKU small project funding (201007176196).

\end{document}